\documentclass[]{article}
\usepackage{caption}
\usepackage{subcaption}
\usepackage{graphicx, xcolor}
\usepackage{textcomp}
\usepackage{pgfplots, url}
\usepackage{pgf-pie}

\usepackage[margin=.65in]{geometry}
\begin{document}

\title{Understanding DNS Query Composition at B-Root}

\author{Jacob Ginesin, \textit{Northeastern University,}
        Jelena Mirkovic,~\textit{USC/ISI}}
        


\date{}
\maketitle

\begin{abstract}

The Domain Name System (DNS) is part of critical internet infrastructure, as DNS is invoked whenever a remote server is accessed (an URL is visited, an API request is made, etc.) by any application. DNS queries are served in hierarchical manner, with most queries served locally from cached data, and a small fraction propagating to the top of the hierarchy -- DNS root name servers. Our research aims to provide a comprehensive, longitudinal characterization of DNS queries received at B-Root over ten years. We sampled and analyzed a 28-billion-query large dataset from the ten annual ``Day in the Life of the Internet (DITL)'' experiments from 2013 through 2022. We sought to identify and quantify unexpected DNS queries, establish longitudinal trends, and compare our findings with published results of others. We found that unexpected query traffic increased from  39.57\% in 2013 to 67.91\% in 2022, with 36.55\% of queries being priming queries. We also observed growth and decline of Chromium-initiated, random DNS queries. Finally, we analyzed the largest DNS query senders and established that most of their traffic consists of unexpected queries.
\end{abstract}


\vspace{-2mm}

\section{Introduction}

The Domain Name System (DNS) is the Internet's system for mapping alphanumeric resource names (e.g. the name of a web or a mail server) to their respective values (in most cases an IP address). This system is critical for basic functioning of the Internet. DNS queries are issued whenever a remote server is accessed by a client application, usually to obtain an IP address for a given server's name. A missing or an incorrect reply to such queries can halt all communication between the server and the client.

In many cases DNS queries are answered by a \textit{caching resolver} on the client's local network, which oftentimes has the full response in its cache. If the local caching resolver does not know the answer to a query, it will interact with other DNS participants to obtain, cache, and return the answer to the client. The DNS system is organized as a hierarchy of DNS name servers, with servers at the higher levels of the hierarchy containing information about servers one level lower than itself. At the top of the hierarchy resides 13 DNS roots. Most DNS queries are answered by lower levels of the DNS hierarchy, but some queries propagate to DNS roots. 

 We consider queries that do not follow DNS server naming conventions or occur too frequently with respect to historal data to be \textit{unexpected queries}. Previous work analyzing DNS traces data has revealed a surprising amount of unexpected queries hitting the root servers \cite{Castro2008, Danzig1992, 2003-wessels-d}. Yet, none of the previous work provides a full characterization of unexpected and expected traffic into disjoint and meaningful categories. Such a classification would help us better understand the root causes and severity of different types of unexpected traffic. Our research aims to develop a comprehensive classification of DNS queries, and use it to study trends in DNS query traffic at B-root over the past ten years


We make the following contributions in this paper: 
\begin{enumerate}
    \item We propose a detailed, comprehensive DNS query classification scheme to cover main root causes of unexpected DNS traffic. 
    \item We quantify unexpected DNS query traffic at B-root, one of 13 DNS roots, both in aggregate and per class of interest. We study longitudinal trends in unexpected DNS queries over the course of ten years, using annually collected ``Day in the Life of the Internet (DITL)'' data~\cite{dnsoarcDITLTraces}. We find an increase in unexpected traffic --- from 39.57\% in 2013 to 67.91\% in 2022. We additionally find 36.55\% of traffic in 2022 is due to empty queries.
    \item We identify top senders of DNS queries to B-root, then classify the traffic coming from each sender. We reveal most traffic from top senders consists of unexpected queries.
\end{enumerate}

\section{Background and Related Work}



In this section we provide more details about DNS hierarchy and query resolution, as well as discuss prior work characterizing traffic at DNS query resolvers.

\subsection{DNS Hierarchy}

DNS queries are issued by applications and operating systems whenever a connection is established with a remote server. For example, if a user types the URL \texttt{www.example.com} into their browser, a DNS query containing the aforementioned name is sent to discover the corresponding IP address. The query is first sent to a \textit{caching resolver} -- usually a server in the same local network as the query sender. The caching resolver (\textit{resolver} for short) will attempt to respond by searching for the query's answer in its cache. If the full answer is not in the cache, the resolver will interact with different \textit{authoritative name servers} to try to determine the full answer. Such an answer will be returned to the client, then saved in cache to respond to potential future queries.

The DNS utilizes a distributed, hierarchical zoning system in order to designate authority, ensure the system's robustness, and effectively distribute query traffic across servers. Each name, e.g. \texttt{www.example.com}, can be viewed as collection of name segments separated by dots \cite{rfc1035}. Each name segment resides in a separate name space, each of which has a designated \textit{authoritative name server}. Such servers will answer queries about names within that name space, either by providing the full answer or by directing the query sender to an authoritative name server for a subset of the given name space. In our example, the caching resolver trying to find the IP address for \texttt{www.example.com} may have in its cache the name and IP address of the name server authoritative for \texttt{.com} \textit{top-level domain (TLD)}. The resolver may repeat its query to the TLD name server, and receive back the name and IP address of the name server authoritative for \texttt{example.com} \textit{second-level domain (sTLD)}. The resolver will cache this new information, then repeat its query to this sTLD name server and receive a full response, which will be cached and returned to the client.

If the resolver from our example does not have information about the relevant TLD server (e.g., \texttt{.com}), it will send its query to one of DNS root name servers. The names and IP addresses of root name servers are often hard-coded in operating system releases, thus a resolver always knows how to reach a DNS root. The root zone is served by 13 logical root name servers (13 root server names, A--M) and hundreds of physical servers.\footnote{The number of root servers used to be capped at 13 due to the size limitation of UDP reply packets. After the introduction and rollout of anycast, the number of physical servers per DNS root has increased.} The root name server will provide the name and IP address of the relevant TLD server, which will be cached by the caching resolver. The rest of name resolution continues as described in the previous paragraph.

Responses from DNS replies carry a ``time-to-live (TTL)'' value, a number of seconds that the authoritative server suggests they should be cached. A resolver can decide to cache a DNS record for a shorter time than the recommended TTL. DNS records for names higher in the hierarchy, such as sTLD and TLD servers, usually have TTL values in hours or even days. Caching should ensure that a resolver can quickly reply to most client DNS queries, and that higher levels of DNS hierarchy do not receive too frequent queries from any resolver.

Two recent extensions to DNS protocol introduce changes to the query resolution process. Query minimization (QMIN) RFC 7816~\cite{rfc7816} instructs DNS resolvers to protect their clients' privacy by only asking each authoritative name server for the name segment that the resolver is currently trying to resolve. In our example, the resolver would not send the full \texttt{www.example.com} query to each authoritative name server. Instead, it would send a \texttt{com} query to a root name server, a \texttt{example.com} query to the TLD name server, and the full query to the sTLD name server. RFC 8109~\cite{rfc8109} introduces \textit{priming queries}, i.e., queries of type nameserver (NS) for the root zone ``.''. Such queries can be sent to any root, and the reply should specify all root server names and IP addresses. Priming queries can help a resolver learn a new IP address for a root name server. In practice, the mapping between root server names and IP addresses has been stable enough as to not require additional root servers to be introduced \cite{ianaRootServers}. 




Finally, while the most popular DNS query maps a DNS name into an IP address (query type A for IPv4 address, query type AAAA for IPv6 address), there are other types of DNS queries. A NS query returns the name and often the IP address of the authoritative name server for the specified query name. A pointer record (PTR) query asks for reverse mapping from an IP address into a DNS name. A start-of-authority (SOA) query requests some metadata about the name, such as the email address of the administrator, when the domain was last updated, and how long the server should wait between refreshes. Mail exchanger (MX) queries are for mail servers serving a given name.  There exists several other, less frequent, query types \cite{rfc1034}.


\subsection{DNS Query Classification}

Castro et al.~\cite{Castro2008} analyzed DITL datasets at eight root servers from years 2006 through 2008. They uncovered very high volumes of unexpected traffic across the root zone --- almost 98\% of queries were identified as unexpected. Unexpected queries were specified to fall into any one of the following categories: invalid query class (a field in a DNS query with five valid values), A or AAAA queries where the query name is an address, queries with invalid TLDs, queries with non printable characters or underscores, PTR queries for a private IP address, identical queries (same class, type, name and ID), repeated queries (same class, type, and name but different ID), and queries where referral records (TLD and sTLD) have not been properly cached. Our study is in a sense a modern sequel to this previous work. We extend the Castro et al. study in a few ways: (1) we dive deeper into invalid query categories, characterizing them by their root cause, (2) our analysis covers newer DNS query trends, like query minimization and priming queries, (3) our analysis spans ten years of DITL data albeit at only one root, and (4) we analyze top senders in several invalid query categories to investigate if there are any commonalities between them that would explain their querying behavior.


\subsection{DNS Sender Analysis}
A recent study measured centralization in senders to B-Root, with a specific focus on tracking the 5 top cloud providers \cite{Moura2020}. This study reveals that in 2020, more than 30\% of all queries to two TLD name servers and B-root were sent from five large cloud providers: Google, Amazon, Microsoft, Facebook, and Cloudflare. We extend upon this work by examining and ranking all senders at B-root instead of just cloud providers. We find senders which often send at rates that are too high often send predominantly queries with invalid TLDs. A similar trend was observed by Castro et al.~\cite{Castro2008} in 2006--2008.


\section{Dataset}

Each year, most root servers and several TLD servers collect and publish all their query traffic on a specific, predetermined day. This effort is known as ``Day in the Life of the Internet'' or DITL, and is undertaken to produce useful data for research \cite{dnsoarcDITLTraces}. Although DITL data has been collected at other root name servers since 2006, B-Root joined the experiment in 2013; likewise, our sample covers just ten years of traffic (2013--2022)~\cite{isiDatasets}. Ideally, we would have analyzed all roots' data from DITL collection. However, this data is only available on OARC servers, which have limited computational power. For this reason, we started with B-root data, which was available locally at our servers, and we plan to extend our analysis to other roots' data in the future.


To speed up our analysis, we analyzed samples from B-Root's DITL data. For each year of DITL experiment data, we utilized the sample function from Python's \texttt{random} package \cite{van1995python} to generate ten year-specific subsets of data. For each of our 10 subsets, we generate 4 additional subsets each denoting one of four time segments:  12-1am, 6-7am, 12-1pm, and 6-7pm. Table \ref{table:1} shows the details of the dataset we analyze in this paper. In total, we study 28 billion DNS queries spread over 10 years, one day per year. 

\begin{table}[h!]
\centering
\begin{tabular}{c c c c c c} 
\textbf{Date (Y/M/D)} & \textbf{Main} & \textbf{12-1am} & \textbf{6-7am} & \textbf{12-1pm} & \textbf{6-7pm} \\
\noalign{\vskip 1mm}    

2013/05/28 & 1.00B & 19.96M & 37.68M & 69.44M & 48.53M\\
\noalign{\vskip .5mm} \hline \noalign{\vskip 1mm}    
2014/04/28 & 0.98B & 255.10M & 61.24M & 11.19M & 46.80M\\ 
\noalign{\vskip .5mm} \hline \noalign{\vskip 1mm}   
2015/04/13 & 0.96B & 35.87M & 42.82M & 63.72M & 29.55M\\
\noalign{\vskip .5mm} \hline \noalign{\vskip 1mm}   
2016/04/05 & 4.22B & 148.62M & 154.42M & 344.51M & 172.56M\\
\noalign{\vskip .5mm} \hline \noalign{\vskip 1mm}   
2017/04/11 & 3.50B & 134.81M & 128.26M & 183.75M & 172.21M\\
\noalign{\vskip .5mm} \hline \noalign{\vskip 1mm}   
2018/04/10 & 3.90B & 97.01M & 121.28M & 255.69M & 181.40M\\
\noalign{\vskip .5mm} \hline \noalign{\vskip 1mm}   
2019/04/08 & 3.63B & 93.38M & 157.30M & 295.90M & 103.52M\\
\noalign{\vskip .5mm} \hline \noalign{\vskip 1mm}   
2020/05/05 & 3.52B & 67.86M & 144.47M & 267.39M & 116.75M\\
\noalign{\vskip .5mm} \hline \noalign{\vskip 1mm}   
2021/04/13 & 2.07B & 79.78M & 69.06M & 112.40M & 93.40M\\
\noalign{\vskip .5mm} \hline \noalign{\vskip 1mm}   
2022/04/12 & 4.11B & 92.40M & 105.70M & 326.66M & 87.48M\\

\end{tabular}
\caption{Evaluated Datasets}
\label{table:1}
\end{table}

\vspace{-2mm}

\section{Methodology}\label{sec:method}

In this section we describe our methodology to determine query classes, and how we implemented our approach.

\subsection{Classification Goals}

While previous works have primarily focused on opportunistically measuring some aspects of unexpected queries, our goal is to provide a more comprehensive, general classification. We seek to define a method to allow us to stratify DNS queries into sections denoting different root causes of unexpected traffic. To do this, we consider two qualities to be of interest when creating our classification method: full-coverage and mutual exclusivity. A method that has full-coverage places every single query of a given dataset into a single, defined category at each level of classification. A method that is mutually exclusive ensures there's no overlap between query categories at the same classification level. 


We opt to primarily classify queries based on query names; yet, we also recognize it's possible to classify queries by other features (e.g. query types, cached/uncached queries, repeated queries) and we hope to do so in the future. In developing our name-based method that fulfills the aforementioned qualities, we consider the DNS zoning hierarchy as defined by RFC 1035 \cite{rfc1035}. In moving down the zoning hierarchy from the root zone (right to left in the context of a textual DNS query), we recognize three mutually exclusive possibilities: the query is empty (``.''), the query ends following some text (e.g. ``foo.''), or the query continues (e.g. ``[more query].foo.''). This method is recursive on the latter query case, which is relevant for processing queries whose subdomains are multi-level \cite{rfc1035}. 

In accordance with our method, we stratified our data into three all-encompassing, mutually exclusive categories - \textit{empty} (``.''), \textit{has-TLD} (e.g. ``example.com''), and \textit{one-word} (e.g. ``foobar.''). 
Next we split the has-TLD category into \textit{valid-TLD} and \textit{invalid-TLD} by comparing the TLD of each query to IANA's maintained valid TLD list \cite{icannListTopLevel}. Within the one-word category we also attempt to detect presence of valid TLDs, which can occur due to query minimization \cite{rfc7816}.
We further stratified the valid-TLD category by categorizing valid TLDs by frequency.

Before categorizing invalid-TLD queries by TLD frequency, we separated classifications of queries we deemed interesting. We quantified queries that contained top-level domains consisting of entirely numbers because they're deemed invalid by RFC1034 \cite{rfc1034}. We quantified queries from Appletalk, a discontinued proprietary suite of networking protocols for Apple products, as it could potentially indicate legacy Apple product usage \cite{archiveV106Printing}, leaking private data into the public Internet. We quantified queries with TLDs containing ``bad encoding'' (ASCII depicted as ``\textbackslash{}xxx\textbackslash{}xxx'') because of its high frequency in DITL data. Because Chromium-initiated queries are known to occasionally contain an invalid-TLD \cite{apnicChromiumsImpact}, we quantified those as well (the importance of Chromium-initiated queries is discussed further in Section \ref{sec:chromium}). 

We separated Chromium-initiated queries from within the one-word category due to their overabundance in certain years. Chromium-initiated queries are discussed further in Section \ref{sec:chromium}. We quantified minimized queries (minimized queries at root servers look like one-word queries whose content is a valid top-level domain) in our collections after the technique's introduction in March of 2016 \cite{rfc7816}. Minimized queries and their importance are discussed further in Section \ref{sec:qmin}.



Our implementation of our classification method involves use of dictionary-based matching and regular expressions. We achieve exclusivity by enforcing the order in which we apply classification criteria within a Python program. 


\section{Results}

In this Section we present our results. We show the breakdown of DNS query traffic in 2013 and 2022 in Section \ref{sec:breakdown}. We analyze trends in query types in Section \ref{sec:types}. We analyze longitudinal trends in Section \ref{sec:trends}. We explore top senders of queries to B-Root in \ref{sec:senders}. We specifically explore Chromium-initiated queries in Section \ref{sec:chromium}. We quantify the increasing presence of minimized queries in Section \ref{sec:qmin}. We explore empty queries in Section \ref{sec:empty}. 


\subsection{2013 \& 2022 B-Root Traffic Breakdown and Comparison}
\label{sec:breakdown}

\begin{figure}[h]
  \centering
  \includegraphics[width=9cm]{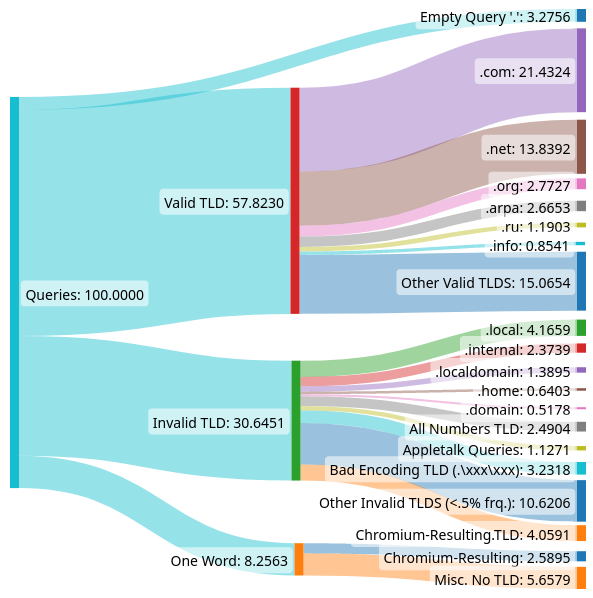}
  \caption{Classification of 1.00 billion DNS traces at B-Root in 2013}
  \label{fig:2013}
\end{figure}

\begin{figure}[h]
  \centering
  \includegraphics[width=9cm]{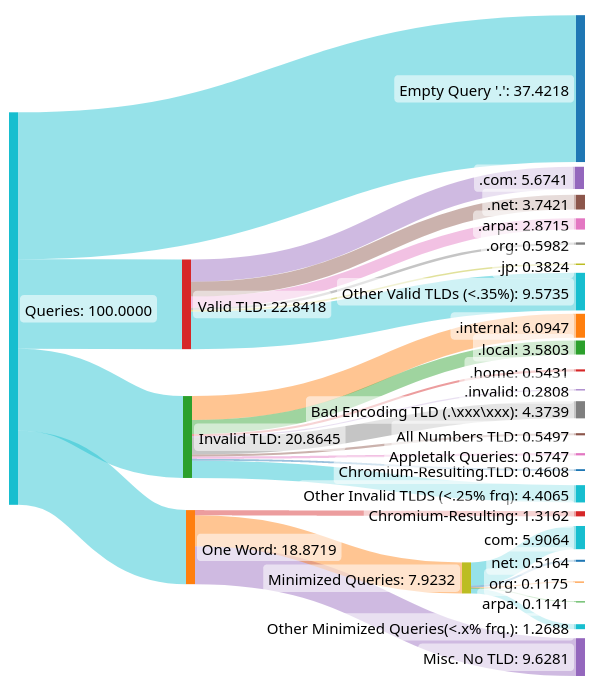}
  \caption{Classification of 4.11 billion DNS traces at B-Root in 2022}
  \label{fig:2022}
\end{figure}

We applied our classification method to DITL collections of B-Root DNS traces from 2013 and 2022 with the intent of revealing trends over the past ten years (see Figures \ref{fig:2013} and \ref{fig:2022} respectively). After splitting our data into the four previously designated categories - empty, valid-TLD, invalid-TLD, and one-word - we did additional work to further stratify each category. Within valid-TLD, we quantified the highest frequency valid top-level domains. Within invalid-TLD, we quantified Appletalk queries \cite{archiveV106Printing}, queries with top-level domains that are incorrectly encoded (e.g. "[query].\textbackslash{}xxx\textbackslash{}xxx\textbackslash{}xxx"), queries with all-number top-level domains\footnote{All-number top-level domains are explicitly specified as invalid by RFC1034.}, and Chromium-initiated queries (see Section \ref{sec:chromium}). Beyond these specific categories within invalid-TLD, we quantified the highest frequency unique invalid top-level domains. Within the One-Word category, we quantified Chromium-resulting queries. Because minimized queries were introduced in 2016, we characterized them only in our 2022 dataset. 


Between 2013 and 2022, we see a 34\% increase in empty queries, a 36\% reduction in valid-TLD queries, a 10\% reduction in invalid-TLD queries, and a 10\% increase in one-word queries. The large increase in empty queries is significant and could be due to priming queries, as discussed in Section \ref{sec:empty}.


Within valid-TLD, we see a sharp reduction in the percentage of \texttt{.com} queries (21.43\% to 5.67\%), \texttt{.net} queries (13.83\% to 3.74\%), and .org queries (2.77\% to 0.60\%). Surprisingly, \texttt{.arpa} queries stay at approximately the same percentage across the 10 year gap (2.66\% to 2.87\%).

Within invalid-TLD, we see a small increase in \texttt{.internal} queries despite an overall reduction in the category---this is potentially indicative of a persistent, growing leak. Appletalk queries decrease from 1.13\% to .57\%, which is expected given Appletalk is long defunct \cite{archiveV106Printing}. 

\subsection{Query Types}\label{sec:types}

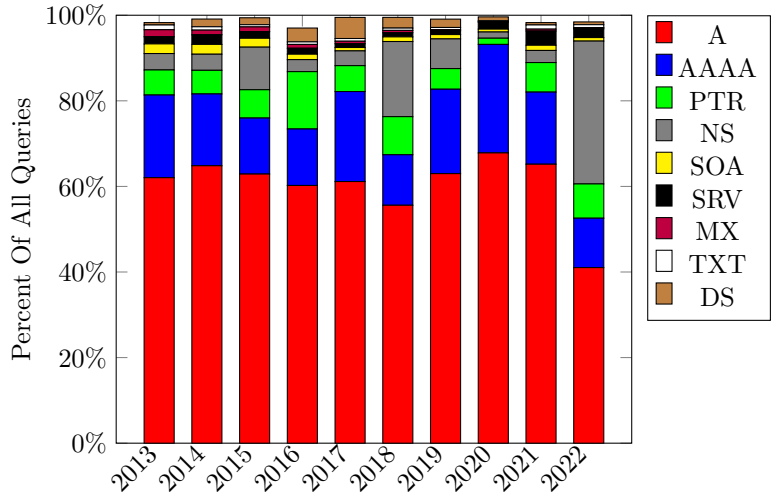
\begin{figure}[]
  \centering
\begin{tikzpicture}
  \begin{axis}[
    ybar stacked, ymin=0,  
    ymax=100,
    bar width=4mm,
    legend pos=outer north east,
    symbolic x coords={
        2013, 
        2014,
        2015,
        2016,
        2017,
        2018,
        2019,
        2020,
        2021,
        2022
    },
    x tick label style={rotate=45,anchor=east},
    ylabel=Percent Of All Queries,
    yticklabel= {\pgfmathprintnumber\tick\%},
    xtick=data,
    nodes near coords align={anchor=north},
    every node near coord/.style={
    },
  ]
  \addplot [fill=red] coordinates {
    ({2013}, 62.04)
    ({2014}, 64.83)
    ({2015}, 62.89)
    ({2016}, 60.22)
    ({2017}, 61.14)
    ({2018}, 55.60)
    ({2019}, 62.99)
    ({2020}, 67.83)
    ({2021}, 65.19)
    ({2022}, 41.03)
  };
  
  \addplot [fill=blue] coordinates {
    ({2013}, 19.32)
    ({2014}, 16.78)
    ({2015}, 13.09)
    ({2016}, 13.20)
    ({2017}, 21.03)
    ({2018}, 11.80)
    ({2019}, 19.72)
    ({2020}, 25.35)
    ({2021}, 16.85)
    ({2022}, 11.54)
  };
  
  \addplot [fill=green] coordinates {
    ({2013}, 5.85)
    ({2014}, 5.52)
    ({2015}, 6.61)
    ({2016}, 13.40)
    ({2017}, 6.05)
    ({2018}, 8.90)
    ({2019}, 4.82)
    ({2020}, 1.45)
    ({2021}, 6.90)
    ({2022}, 8.01)
  };
  
  \addplot [fill=gray] coordinates {
    ({2013}, 3.84)
    ({2014}, 3.76)
    ({2015}, 9.95)
    ({2016}, 2.81)
    ({2017}, 3.45)
    ({2018}, 17.56)
    ({2019}, 6.94)
    ({2020}, 1.45)
    ({2021}, 2.83)
    ({2022}, 33.38)
  };

  \addplot [fill=yellow] coordinates {
    ({2013}, 2.25)
    ({2014}, 2.30)
    ({2015}, 2.08)
    ({2016}, 1.27)
    ({2017}, 0.79)
    ({2018}, 1.08)
    ({2019}, 1.01)
    ({2020}, 0.677)
    ({2021}, 1.22)
    ({2022}, 0.82)
  };

    \addplot [fill=black] coordinates {
    ({2013}, 1.68)
    ({2014}, 2.25)
    ({2015}, 1.56)
    ({2016}, 1.39)
    ({2017}, 0.89)
    ({2018}, .97)
    ({2019}, .89)
    ({2020}, 1.60)
    ({2021}, 3.33)
    ({2022}, 2.07)
  };

    \addplot [fill=purple] coordinates {
    ({2013}, 1.63)
    ({2014}, 1.08)
    ({2015}, 1.09)
    ({2016}, .89)
    ({2017}, .58)
    ({2018}, .55)
    ({2019}, .26)
    ({2020}, .16)
    ({2021}, .43)
    ({2022}, .27)
  };

    \addplot [fill=white] coordinates {
    ({2013}, 1.11)
    ({2014}, .78)
    ({2015}, .51)
    ({2016}, .59)
    ({2017}, .63)
    ({2018}, .54)
    ({2019}, .51)
    ({2020}, .20)
    ({2021}, .98)
    ({2022}, .67)
  };

    \addplot [fill=brown] coordinates {
    ({2013}, .56)
    ({2014}, 1.81)
    ({2015}, 1.63)
    ({2016}, 3.24)
    ({2017}, 4.93)
    ({2018}, 2.50)
    ({2019}, 1.95)
    ({2020}, 0.90)
    ({2021}, 0.55)
    ({2022}, 0.65)
  };

  \legend{A, AAAA, PTR, NS, SOA, SRV, MX, TXT, DS}
  \end{axis}
\end{tikzpicture}
  \caption{Breakdown of DNS query types at B-Root from 2013 through 2022}
  \label{fig:type}
\end{figure}

Figure \ref{fig:type} shows the distribution of queries by type at B-Root from years 2013 through 2022. For all years, A-Type queries, used to request an IPv4 address for a given query name, are the most common (60\% of the total previous to 2022). AAAA-type queries, used to request IPv6 an address, are generally the second most common query type (15\% of the total previous to 2022). In 2022, we measure a large reduction in A and AAAA-type queries and a large increase in NS-type queries. The increase in NS-type queries is associated with the implementation of resolver priming \cite{rfc8109}---priming queries are further discussed in Section \ref{sec:qmin}.

\vspace{-4mm}

\subsection{Longitudinal Trends}\label{sec:trends}

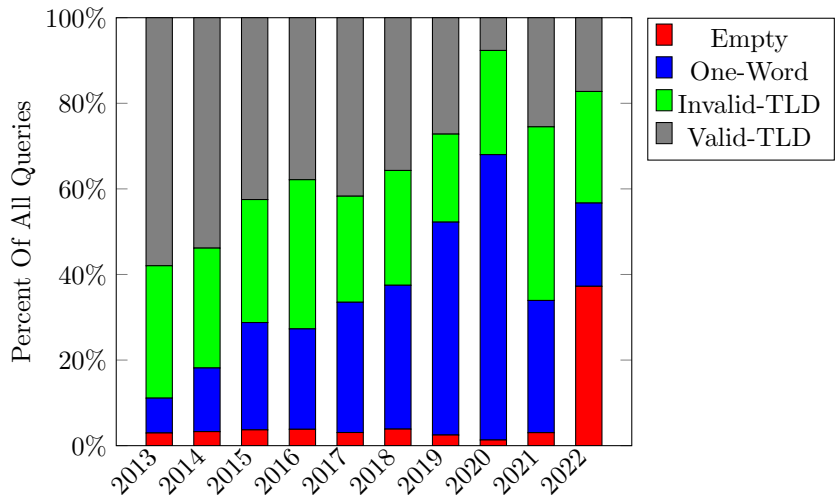
\begin{figure}[]
  \centering
\begin{tikzpicture}
  \begin{axis}[
    ybar stacked, ymin=0,  
    ymax=100,
    bar width=3.5mm,
    legend pos=outer north east,
    symbolic x coords={
        2013, 
        2014,
        2015,
        2016,
        2017,
        2018,
        2019,
        2020,
        2021,
        2022
    },
    x tick label style={rotate=45,anchor=east},
    ylabel=Percent Of All Queries,
    yticklabel= {\pgfmathprintnumber\tick\%},
    xtick=data,
    nodes near coords align={anchor=north},
    every node near coord/.style={
    },
  ]
  \addplot [fill=red] coordinates {
    ({2013}, 2.9603)
    ({2014}, 3.2808)
    ({2015}, 3.6901)
    ({2016}, 3.8242)
    ({2017}, 3.0692)
    ({2018}, 3.8709)
    ({2019}, 2.4860)
    ({2020}, 1.3211)
    ({2021}, 3.0366)
    ({2022}, 37.2394)
  };
  
  \addplot [fill=blue] coordinates {
    ({2013}, 8.1450)
    ({2014}, 14.8740)
    ({2015}, 25.0792)
    ({2016}, 23.4651)
    ({2017}, 30.4401)
    ({2018}, 33.6103)
    ({2019}, 49.7478)
    ({2020}, 66.6825)
    ({2021}, 30.8924)
    ({2022}, 19.4543)
  };
  
  \addplot [fill=green] coordinates {
    ({2013}, 30.9352)
    ({2014}, 28.0068)
    ({2015}, 28.7403)
    ({2016}, 34.8556)
    ({2017}, 24.7834)
    ({2018}, 26.8245)
    ({2019}, 20.5680)
    ({2020}, 24.3431)
    ({2021}, 40.5737)
    ({2022}, 26.0706)
  };
  
  \addplot [fill=gray] coordinates {
    ({2013}, 57.9595)
    ({2014}, 53.8383)
    ({2015}, 42.4903)
    ({2016}, 37.8551)
    ({2017}, 41.7073)
    ({2018}, 35.69440)
    ({2019}, 27.1982)
    ({2020}, 7.6533)
    ({2021}, 25.4973)
    ({2022}, 17.2357)
  };

  \legend{Empty, One-Word, Invalid-TLD, Valid-TLD }
  \end{axis}
\end{tikzpicture}
  \caption{Breakdown of longitudinal trends from 2013 through 2022 in DITL datasets at B-Root}
  \label{fig:trends}
\end{figure}

We applied our classification method to each of our collections of B-Root DNS traces from 2013 through 2022 with the intent of discovering longitudinal trends. Figure \ref{fig:trends} shows the breakdown of empty, one-word, invalid-TLD, and valid-TLD queries for each year 2013--2022. Valid-TLD queries consistently decline from 57.82\% in 2013 to 22.84\% in 2022. invalid-TLD queries stay approximately constant through the 10 year sample, hovering between 20\% and 30\% of all queries. One-word queries see a steady increase from 8.26\% in 2013 to 68.45\% in 2020, followed by a sharp decline to 18.87\% in 2020. This rise and fall is largely due to Chromium-resulting queries, as further discussed in section \ref{sec:chromium}. Empty queries hovered around ~3\% until jumping to 37.42\% in 2022. This sudden increase is thought to be a result of excessive priming queries, as discussed in section \ref{sec:empty}.




\vspace{-4mm}

\subsection{Top Senders}\label{sec:senders}

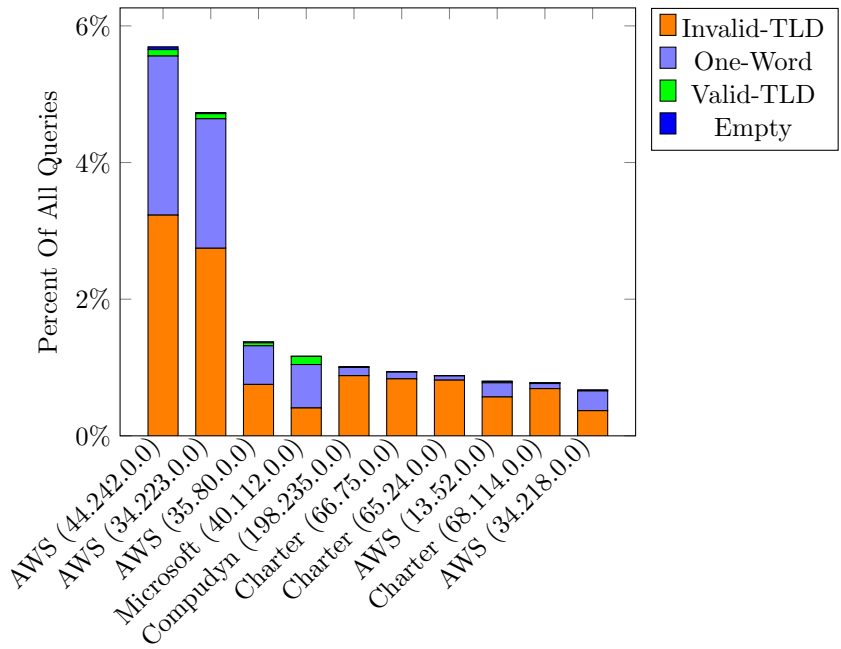
\begin{figure}[]
  \centering
\begin{tikzpicture}
  \begin{axis}[
    yticklabel= {\pgfmathprintnumber\tick\%},
    ybar stacked, ymin=0,  
    legend pos=outer north east,
    bar width=4mm,
    symbolic x coords={
    AWS (44.242.0.0), 
    AWS (34.223.0.0), 
    AWS (35.80.0.0), 
    Microsoft (40.112.0.0), 
    Compudyn (198.235.0.0),
    Charter (66.75.0.0),
    Charter (65.24.0.0),
    AWS (13.52.0.0),
    Charter (68.114.0.0),
    AWS (34.218.0.0)
    },
    x tick label style={rotate=45,anchor=east},
    xtick=data,
    ylabel style={yshift=-0.3cm}, 
    ylabel=Percent Of All Queries,
    nodes near coords align={anchor=north},
    every node near coord/.style={
    },
  ]
  
  \addplot [fill=orange] coordinates {
    ({AWS (44.242.0.0)}, 3.2304)
    ({AWS (34.223.0.0)}, 2.7455)
    ({AWS (35.80.0.0)}, 0.7526)
    ({Microsoft (40.112.0.0)}, 0.4080)
    ({Compudyn (198.235.0.0)},0.8808)
    ({Charter (66.75.0.0)},0.8339)
    ({Charter (65.24.0.0)},0.8153)
    ({AWS (13.52.0.0)},0.5690)
    ({Charter (68.114.0.0)},0.6908)
    ({AWS (34.218.0.0)},0.3684)
  };
  
  \addplot [fill=blue!50] coordinates {
    ({AWS (44.242.0.0)}, 2.3282)
    ({AWS (34.223.0.0)}, 1.8949)
    ({AWS (35.80.0.0)}, 0.5652)
    ({Microsoft (40.112.0.0)}, 0.6356)
    ({Compudyn (198.235.0.0)},0.12038)
    ({Charter (66.75.0.0)},0.0986)
    ({Charter (65.24.0.0)},0.0615)
    ({AWS (13.52.0.0)},0.2071)
    ({Charter (68.114.0.0)},0.0746)
    ({AWS (34.218.0.0)},0.2885)
  };
  
  \addplot [fill=green] coordinates {
    ({AWS (44.242.0.0)}, 0.0942)
    ({AWS (34.223.0.0)}, 0.0746)
    ({AWS (35.80.0.0)}, 0.0434)
    ({Microsoft (40.112.0.0)}, 0.1192)
    ({Compudyn (198.235.0.0)},0.0080)
    ({Charter (66.75.0.0)},0.0049)
    ({Charter (65.24.0.0)},0.0045)
    ({AWS (13.52.0.0)},0.0201)
    ({Charter (68.114.0.0)},0.0120)
    ({AWS (34.218.0.0)}, 0.0100)
  };
    
  \addplot [fill=blue] coordinates {
    ({AWS (44.242.0.0)}, 0.0411)
    ({AWS (34.223.0.0)}, 0.0168)
    ({AWS (35.80.0.0)}, 0.0161)
    ({Microsoft (40.112.0.0)},0.0013)
    ({Compudyn (198.235.0.0)},0.0003)
    ({Charter (66.75.0.0)},0.0011)
    ({Charter (65.24.0.0)},0.0004)
    ({AWS (13.52.0.0)},0.0036)
    ({Charter (68.114.0.0)},0.0015)
    ({AWS (34.218.0.0)},0.0030)
  };

  \legend{Invalid-TLD, One-Word, Valid-TLD, Empty}
  \end{axis}
\end{tikzpicture}
\centering
  \caption{Top query senders in 2022 DITL dataset at B-root}
  \label{figure:senders}
\end{figure}

We identified resolvers that are top senders in DITL dataset from 2022, and show them and their query composition in Figure \ref{figure:senders}. Amazon Web Services (AWS) accounts for the 1st, 2nd, 3rd, 8th, and 10th highest IP host groups and account for approximately ~14\% of all queries to B-Root. AWS sends almost entirely invalid-TLD and one-word queries to B-Root. Microsoft Azure, another cloud computing platform, has a similar query classification breakdown to AWS. This is potentially indicative of rented cloud machines being misconfigured or used for malicious purposes. Charter and Compudyn, both internet service providers, account for ~3.43\% of all traffic to B-Root. Both providers primarily send invalid-TLD queries, potentially indicating a misconfiguration. Additionally, empty and valid-TLD queries aren't present in significant quantities from these large senders. 

\subsection{Chromium-Resulting Queries}\label{sec:chromium}

Chromium is an open-source web browser project primarily maintained by Google. In addition to Google Chrome, several other major web browsers including Microsoft Edge, Opera, Brave, Samsung Internet, and Amazon Silk are based on the Chromium codebase. In total, approximately 75\% of the web-browser market share is Chromium-based \cite{w3counterW3CounterBrowser}.

Chromium includes a feature titled Omnibox, which allows users to enter website names, URLs, or search terms. Chromium then decides if the entered term is a URL or a search term by performing a DNS query. A URL will result in a valid response, while a search term will not --- Chromium can then supply search results from Google. However, a user's machine may be behind a captive portal (e.g., in a hotel), which intercepts each DNS query and responds with either the correct response (e.g., in the URL case) or with a redirect to an internal Web site (e.g., in the search term case). This situation would interfere with the Chromium's response to user input. For this reason each Chromium browser attempts to detect presence of captive portals by sending three randomly generated query names~\cite{chromiumOmniboxHistory} \cite{chromiumNetworkStack}. These queries contain 7--15, lowercase alphabetic characters (e.g., ``daozjwend.''). 

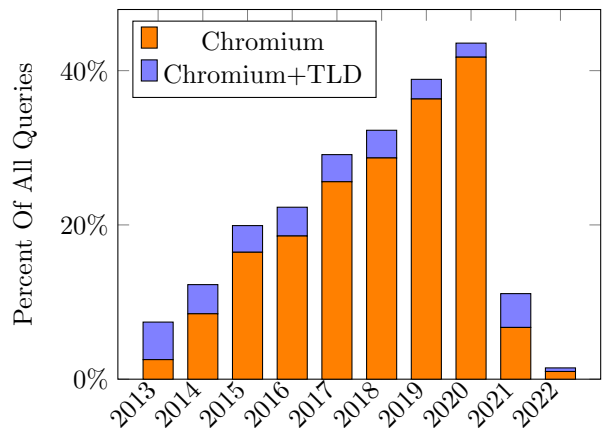
\begin{figure}
  \centering
\begin{tikzpicture}
  \begin{axis}[
    height=6.5cm,
    yticklabel= {\pgfmathprintnumber\tick\%},
    ylabel=Percent Of All Queries,
    width=8cm,
    legend pos=outer north east,
    ybar stacked, ymin=0,
    bar width=4mm,
    legend pos=north west,
    symbolic x coords={
    2013,
    2014,
    2015,
    2016,
    2017,
    2018,
    2019,
    2020,
    2021,
    2022
    },
    x tick label style={rotate=45,anchor=east},
    xtick=data,
    nodes near coords align={anchor=north},
    every node near coord/.style={
    },
  ]
  \addplot [fill=orange] coordinates {
    ({2013}, 2.5354)
    ({2014},8.4787)
    ({2015}, 16.4564)
    ({2016}, 18.5619)
    ({2017}, 25.5965)
    ({2018}, 28.6797)
    ({2019}, 36.3254)
    ({2020}, 41.7465)
    ({2021}, 6.7086)
    ({2022}, 0.9984)
  };

  \addplot [fill=blue!50] coordinates {
    ({2013}, 4.8676)
    ({2014}, 3.7813)
    ({2015}, 3.4557)
    ({2016}, 3.7329)
    ({2017}, 3.5171)
    ({2018}, 3.5923)
    ({2019}, 2.5479)
    ({2020}, 1.8229)
    ({2021}, 4.3845)
    ({2022}, 0.4647)
  };

  \legend{Chromium, Chromium+TLD}
  \end{axis}
\end{tikzpicture}
  \caption{Chromium-initiated queries from 2013 through 2022 in DITL datasets at B-Root}
  \label{fig:chromium}
\end{figure}

As a consequence of this feature combined with Chromium's high market share, root zone name servers have reported a very high quantity of Chromium-originating queries. Our findings at B-Root agree with the findings of previous work quantifying these queries~\cite{apnicChromiumsImpact}. We see a gradual increase in Chromium-originating queries from 2013 through 2020, followed by a sharp decline after 2020 following a change to Omnibox's probing process~\cite{verisignChromiumsReduction}. This trend is shown in Figure \ref{fig:chromium}. Because Chromium-resulting queries have been known to appear both with and without a TLD \cite{apnicChromiumsImpact}, we quantify both types.


\subsection{DNS Query Name Minimization}\label{sec:qmin}

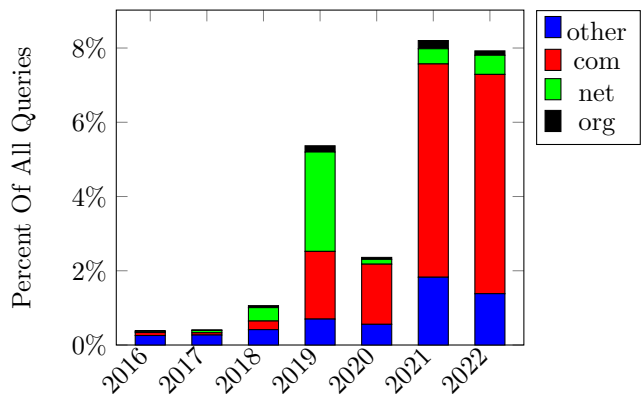
\begin{figure}
  \centering
\begin{tikzpicture}
  \begin{axis}[
    width=7cm,
    ybar stacked, ymin=0,  
    bar width=4mm,
    legend pos=outer north east,
    symbolic x coords={
        2016,
        2017,
        2018,
        2019,
        2020,
        2021,
        2022
    },
    x tick label style={rotate=45,anchor=east},
    ylabel=Percent Of All Queries,
    yticklabel= {\pgfmathprintnumber\tick\%},
    xtick=data,
    nodes near coords align={anchor=north},
    every node near coord/.style={
    },
  ]

  \addplot [fill=blue] coordinates {
    ({2016}, 0.2561)
    ({2017}, 0.2692)
    ({2018}, 0.414)
    ({2019}, 0.6988)
    ({2020}, 0.5574)
    ({2021}, 1.8259)
    ({2022}, 1.3829)
  };

  \addplot [fill=red] coordinates {
    ({2016}, 0.0795)
    ({2017}, 0.0612)
    ({2018}, 0.2319)
    ({2019}, 1.8229)
    ({2020}, 1.6202)
    ({2021}, 5.7452)
    ({2022}, 5.9064)
  };
  
  \addplot [fill=green] coordinates {
    ({2016}, 0.0319)
    ({2017}, 0.0612)
    ({2018}, 0.3611)
    ({2019}, 2.6770)
    ({2020}, 0.1299)
    ({2021}, 0.4091)
    ({2022}, 0.5164)
  };
  
  \addplot [fill=black] coordinates {
    ({2016}, 0.0186)
    ({2017}, 0.0163)
    ({2018}, 0.0536)
    ({2019}, 0.1682)
    ({2020}, 0.0527)
    ({2021}, 0.2213)
    ({2022}, 0.1175)
  };

  \legend{other, com, net, org}
  \end{axis}
\end{tikzpicture}
    \caption{Breakdown of minimized DNS queries at B-Root from 2016 through 2022}
  \label{fig:qmin}
\end{figure}

We seek to quantify the presence of minimized queries (these appear at B-Root as one word valid TLDs, e.g. \texttt{com}) since the inception of the query minimization specification in March of 2016 \cite{rfc7816}. Because the DNS is highly distributed and controlled by hundreds of different organizations, a change in the DNS protocol takes time to propagate. Thus, studying the quantity of minimized queries hitting the root zone could provide insight into the speed at which DNS protocol changes propagate throughout the entire DNS network, potentially aiding in future DNS specification development. In our data, we find a generally steady increase in minimized queries after 2016, as shown in Figure \ref{fig:qmin}. In accordance with the frequency of valid TLDs found in Section \ref{sec:breakdown}, the distribution of minimized queries is expected.


\subsection{Empty Queries} \label{sec:empty}
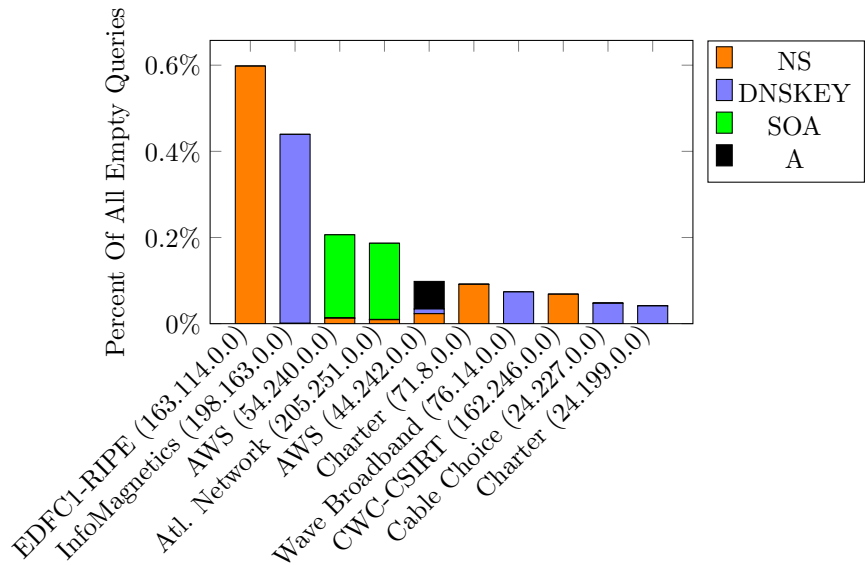
\begin{figure}
  \centering
\begin{tikzpicture}
  \begin{axis}[
    height=5.35cm,
    yticklabel= {\pgfmathprintnumber\tick\%},
    ylabel=Percent Of All Empty Queries,
    width=8cm,
    ybar stacked, ymin=0,  
    legend pos=outer north east,
    bar width=4mm,
    symbolic x coords={
    EDFC1-RIPE (163.114.0.0),
    InfoMagnetics (198.163.0.0),
    AWS (54.240.0.0),
    Atl. Network (205.251.0.0),
    AWS (44.242.0.0),
    Charter (71.8.0.0),
    Wave Broadband (76.14.0.0),
    CWC-CSIRT (162.246.0.0),
    Cable Choice (24.227.0.0),
    Charter (24.199.0.0),
    },
    x tick label style={rotate=45,anchor=east},
    xtick=data,
    nodes near coords align={anchor=north},
    every node near coord/.style={
    },
  ]
  \addplot [fill=orange] coordinates {
    ({EDFC1-RIPE (163.114.0.0)}, 0.5977)
    ({InfoMagnetics (198.163.0.0)}, 0.0009)
    ({AWS (54.240.0.0)}, 0.0127)
    ({Atl. Network (205.251.0.0)}, 0.0092)
    ({AWS (44.242.0.0)}, 0.0232)
    ({Charter (71.8.0.0)}, 0.0913) 
    ({Wave Broadband (76.14.0.0)},0)
    ({CWC-CSIRT (162.246.0.0)}, 0.0683)
    ({Cable Choice (24.227.0.0)}, 0.0003)
    ({Charter (24.199.0.0)}, 0.0003)
  };

  \addplot [fill=blue!50] coordinates {
    ({EDFC1-RIPE (163.114.0.0)}, 0)
    ({InfoMagnetics (198.163.0.0)}, 0.4385)
    ({AWS (54.240.0.0)}, 0.0005)
    ({Atl. Network (205.251.0.0)}, 0.0006)
    ({AWS (44.242.0.0)}, 0.0109)
    ({Charter (71.8.0.0)}, 0)
    ({Wave Broadband (76.14.0.0)}, 0.0737)
    ({CWC-CSIRT (162.246.0.0)}, 0)
    ({Cable Choice (24.227.0.0)}, 0.0470)
    ({Charter (24.199.0.0)}, 0.0409)
  };

  \addplot [fill=green] coordinates {
    ({EDFC1-RIPE (163.114.0.0)}, 0)
    ({InfoMagnetics (198.163.0.0)}, 0)
    ({AWS (54.240.0.0)}, 0.1931)
    ({Atl. Network (205.251.0.0)}, 0.1769)
    ({AWS (44.242.0.0)}, 0)
    ({Charter (71.8.0.0)}, 0.0001)
    ({Wave Broadband (76.14.0.0)}, 0)
    ({CWC-CSIRT (162.246.0.0)}, 0)
    ({Cable Choice (24.227.0.0)}, 0.0007)
    ({Charter (24.199.0.0)}, 0.0001)
  };

  \addplot [fill=] coordinates {
    ({EDFC1-RIPE (163.114.0.0)}, 0)
    ({InfoMagnetics (198.163.0.0)}, 0)
    ({AWS (54.240.0.0)}, 0)
    ({Atl. Network (205.251.0.0)}, 0)
    ({AWS (44.242.0.0)}, 0.0640)
    ({Charter (71.8.0.0)}, 0)
    ({Wave Broadband (76.14.0.0)}, 0)
    ({CWC-CSIRT (162.246.0.0)}, 0)
    ({Cable Choice (24.227.0.0)}, 0)
    ({Charter (24.199.0.0)}, 0)
  };

  \legend{NS, DNSKEY, SOA, A}
  \end{axis}
\end{tikzpicture}
  \caption{Top senders of empty queries in 2022 DITL dataset at B-Root}
  \label{fig:empty}
\end{figure}

One of the most notable outliers we discovered in our data is the overabundance of empty queries in 2022---37.42\% of queries in 2022 to B-Root are empty. Before 2022, empty queries only ever occupied as much as 4\% of all traces. Upon investigation, we find empty query top senders only account for a small fraction of all empty queries, as shwon in Figure \ref{fig:empty}. Similarly, we find each sender, on average, sends 2.8 empty queries to B-Root. We also find 97.20\% of empty queries sent to B-Root in 2022 are type NS. The decentralization and query type indicate the majority of the empty queries hitting B-Root in 2022 are \textit{Priming Queries}, a standard introduced in March of 2017 \cite{rfc8109}. In comparison with the gradual growth of minized query traffic after the introduction of RFC7816, as shown in Figure \ref{fig:trends} the growth of empty queries was nonexistent, then sudden. To the best of our knowledge we are the first to identify this pattern, and in the future we hope to track down the root cause.


\section{Conclusion and Further Directions}

This investigation into B-Root's DNS traces collected from the annual DITL experiment over ten years characterized longitudinal trends, as well as modern issues, such as a high volume of priming queries. Future work involves characterizing valid TLD traffic at B-root and identifying unexpected queries in that category. We would also like to analyze other root's TLD data and see if trends identified at B-root apply to other roots. We encourage other DNS operators to implement our classification method. We also hope to extend our classification approach with more categories in the future. 

\section{Acknowledgement}
This work was performed as a part of a Research Experience for Undergraduates (REU) program, supported by National Science Foundation (NSF) grant \#2051101.



\bibliographystyle{plain} 
{\footnotesize 
\bibliography{refs}}



\section{Appendix}

\subsection{Code}
The code used for processing B-Root's DITL data is publicly available at\\ \url{https://github.com/STEELISI/DITL-Analysis-Scripts}

\subsection{2013-2022 Traffic Breakdown}%
In this section we provide additional diagrams resulting from the application of the classification method described in Section \ref{sec:method} to all 10 years of DITL data. Note the addition of the "minimized queries" category from 2016 and forward.

\begin{figure}[h]
  \centering
  \includegraphics[width=11cm]{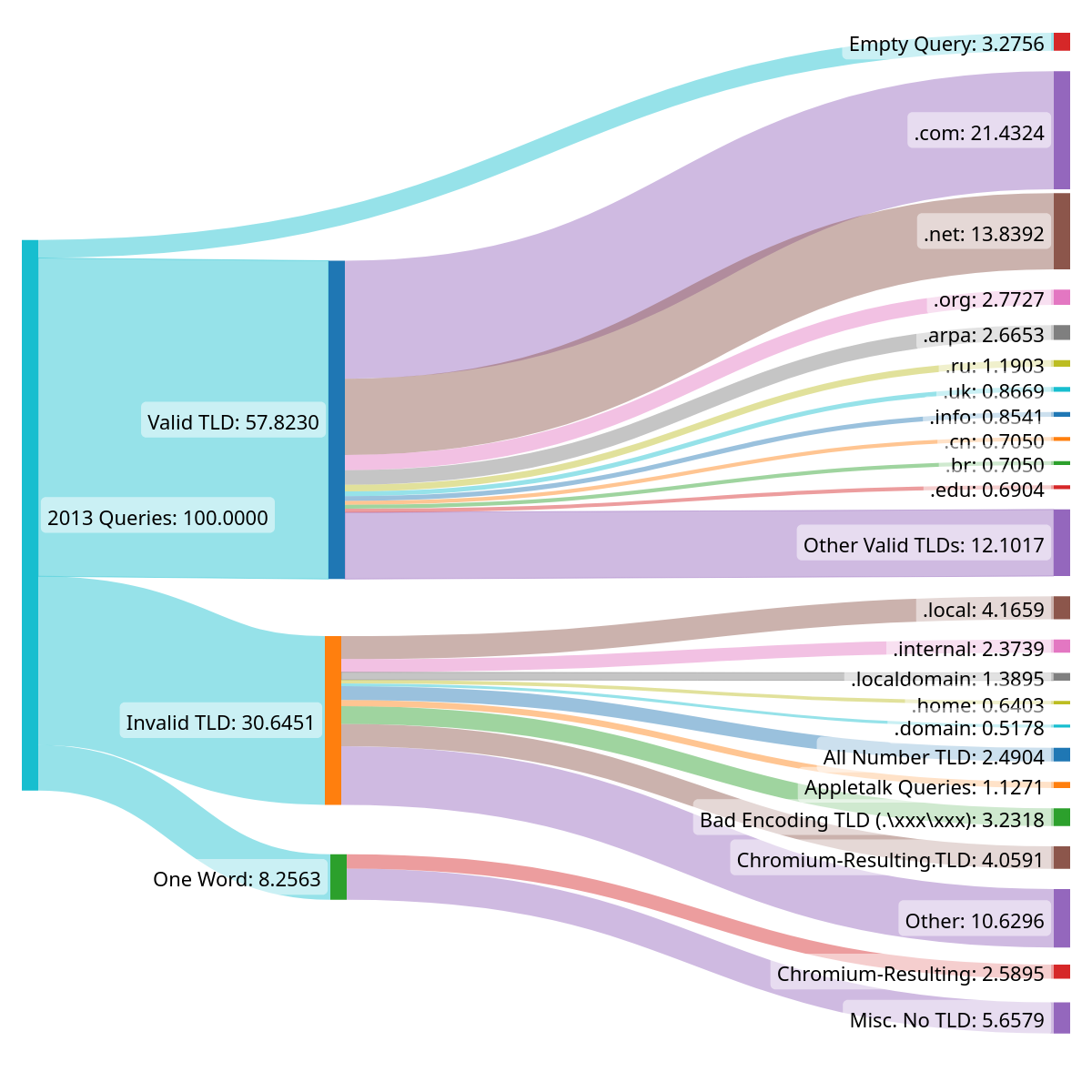}
  \caption{Classification of 1.00 billion DNS traces at B-Root in 2013.}
  \label{fig:2013_appendix}
\end{figure}

\begin{figure}[]
  \centering
  \includegraphics[width=11cm]{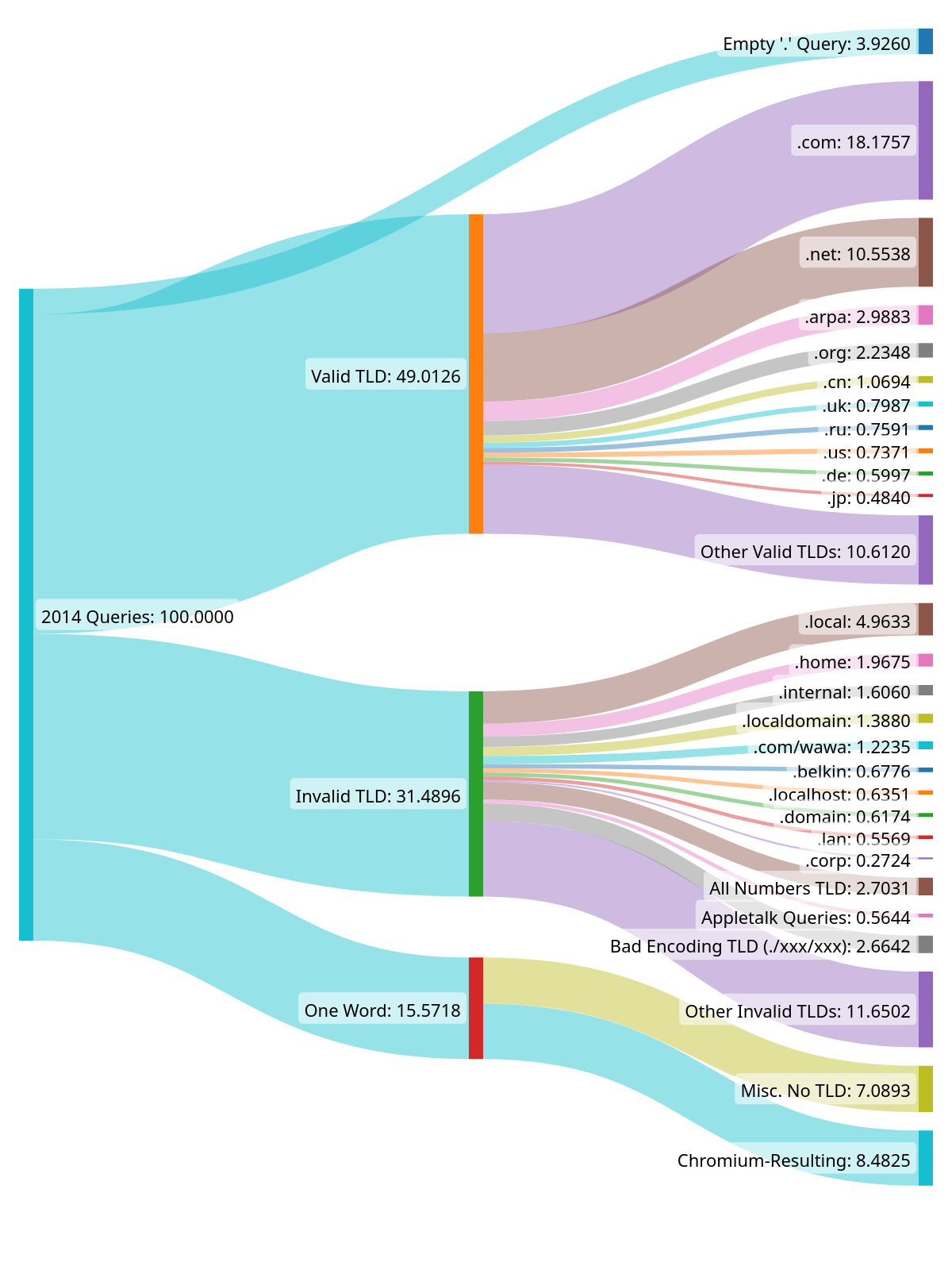}
  \caption{Classification of 0.98 billion DNS traces at B-Root in 2014.}
  \label{fig:2014_appendix}
\end{figure}

\begin{figure}[]
  \centering
  \includegraphics[width=11cm]{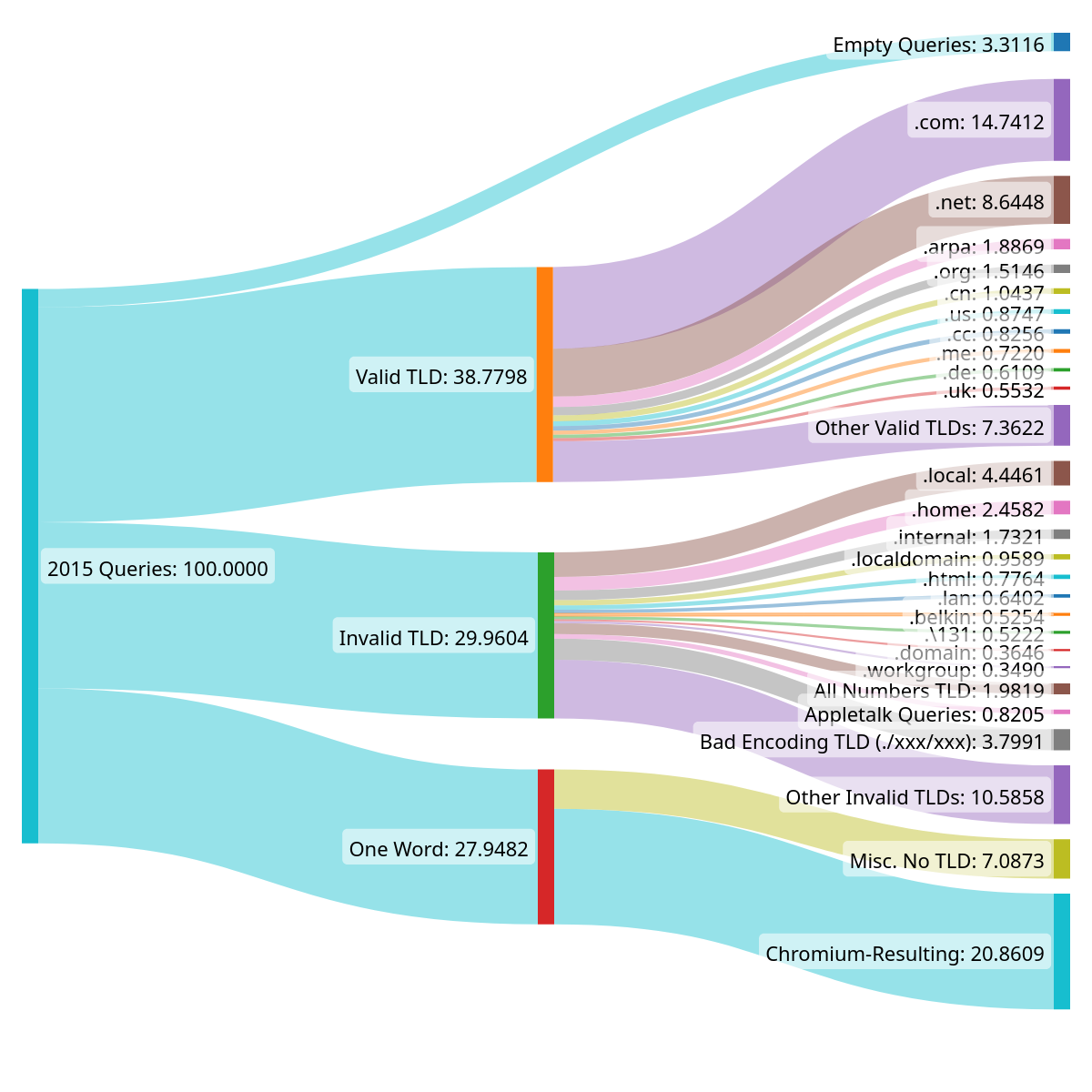}
  \caption{Classification of 0.96 billion DNS traces at B-Root in 2015.}
  \label{fig:2015_appendix}
\end{figure}

\begin{figure}[]
  \centering
  \includegraphics[width=11cm]{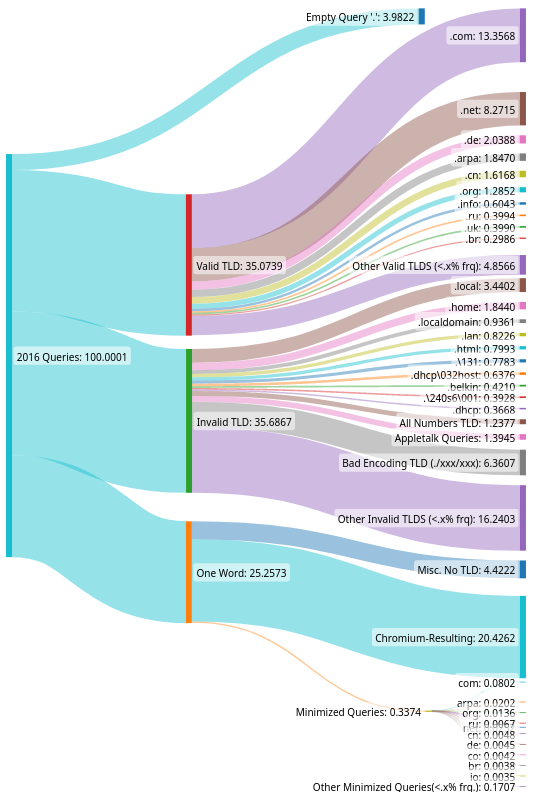}
  \caption{Classification of 4.22 billion DNS traces at B-Root in 2016. }
  \label{fig:2016_appendix}
\end{figure}

\begin{figure}[]
  \centering
  \includegraphics[width=11cm]{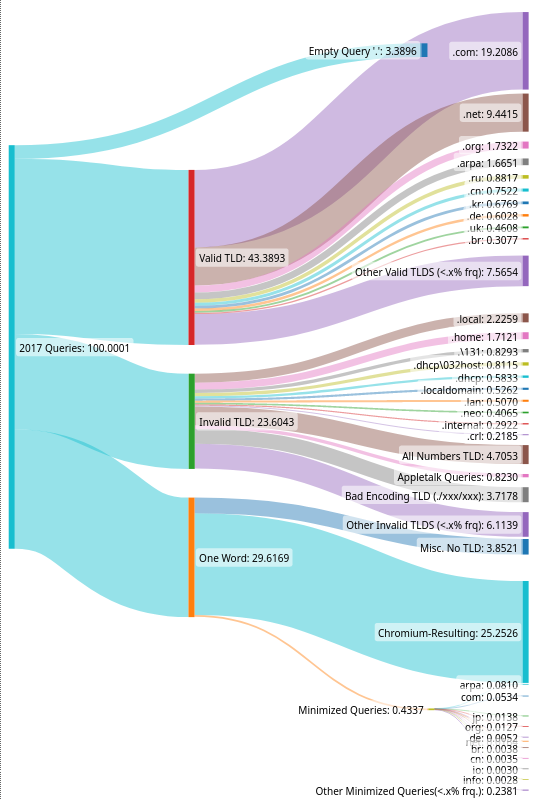}
  \caption{Classification of 3.50 billion DNS traces at B-Root in 2017.}
  \label{fig:2017_appendix}
\end{figure}

\begin{figure}[]
  \centering
  \includegraphics[width=11cm]{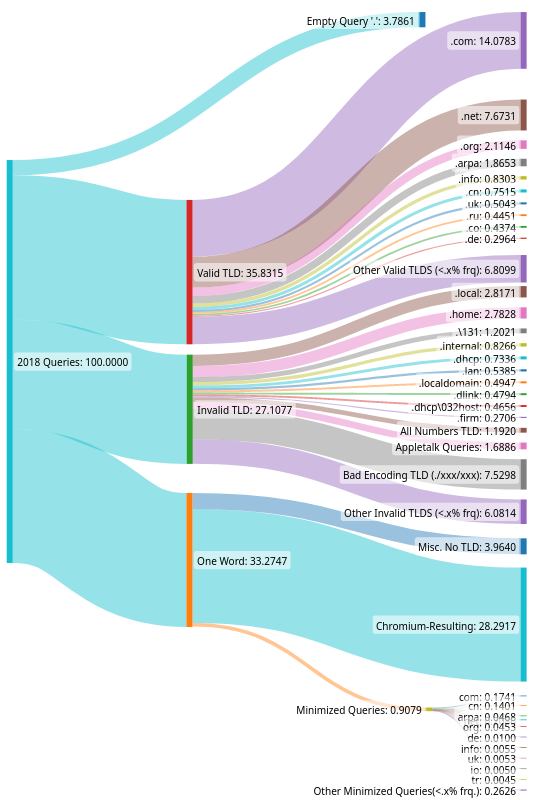}
  \caption{Classification of 3.90 billion DNS traces at B-Root in 2018.}
  \label{fig:2018_appendix}
\end{figure}

\begin{figure}[]
  \centering
  \includegraphics[width=11cm]{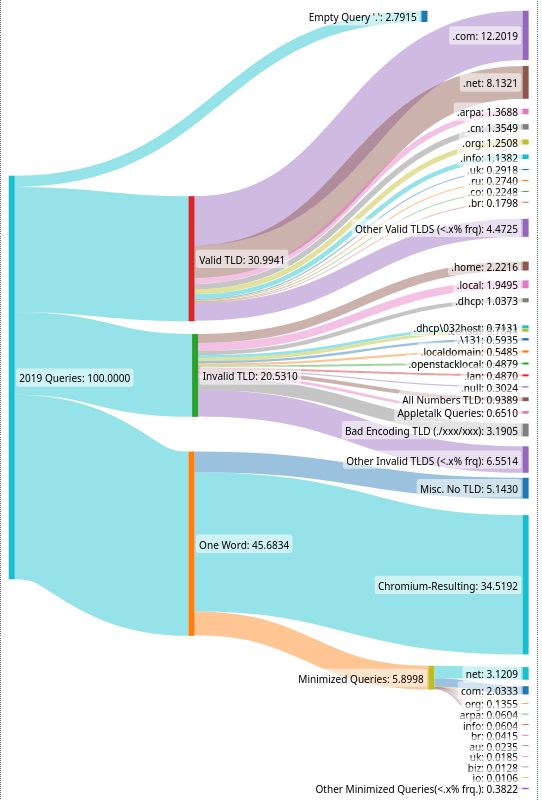}
  \caption{Classification of 3.63 billion DNS traces at B-Root in 2019.}
  \label{fig:2019_appendix}
\end{figure}

\begin{figure}[]
  \centering
  \includegraphics[width=11cm]{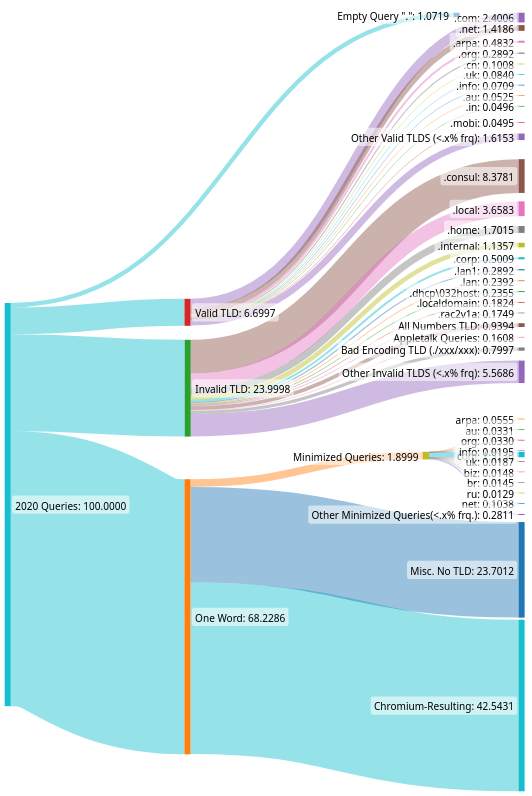}
  \caption{Classification of 3.52 billion DNS traces at B-Root in 2020.}
  \label{fig:2020_appendix}
\end{figure}

\begin{figure}[]
  \centering
  \includegraphics[width=11cm]{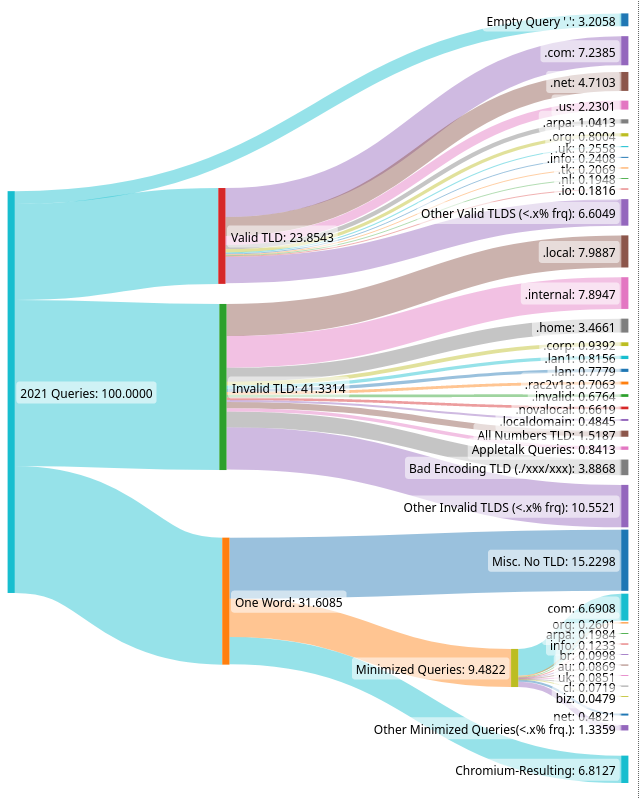}
  \caption{Classification of 2.07 billion DNS traces at B-Root in 2021.}
  \label{fig:2021_appendix}
\end{figure}

\begin{figure}[]
  \centering
  \includegraphics[width=11cm]{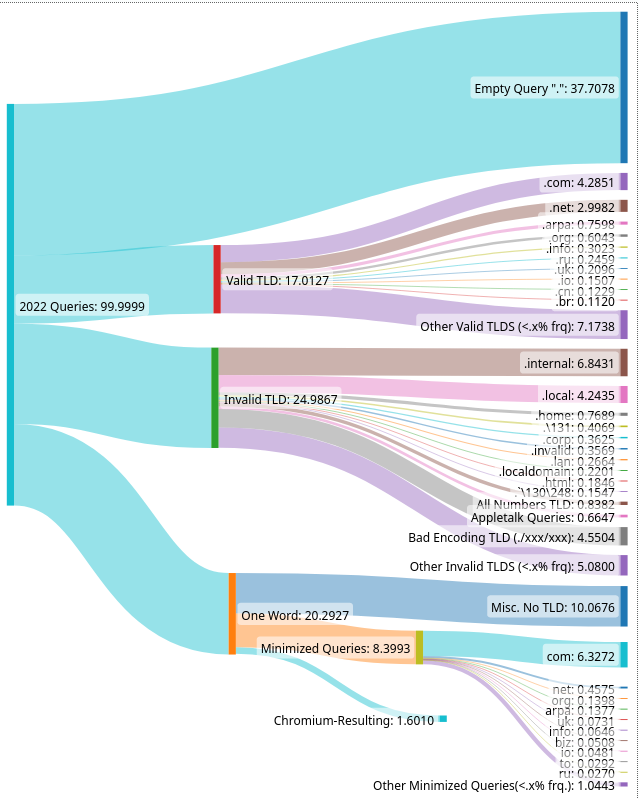}
  \caption{Classification of 100 million DNS traces at B-Root in 2022.}
  \label{fig:2022_appendix}
\end{figure}

\end{document}